\documentclass[aps,prl,showpacs,twocolumn,superscriptaddress]{revtex4}
\usepackage{bbm,bm}
\usepackage{mathrsfs}
\usepackage{epsfig,psfrag}
\usepackage{amsmath,amsfonts,amssymb}
\usepackage[usenames]{color}

\begin{document}

\preprint{draft}

\newcommand{\tec}[1]{{\color{magenta} \{\small \sc #1\}}}
\newcommand{\tei}[1]{{\color{magenta} #1}}

\newcommand{\bx}{{\bf x}}
\newcommand{\bz}{{\bf z}}
\newcommand{\bq}{{\bf q}}
\newcommand{\bu}{{\bf u}}
\newcommand{\bn}{{\bf n}}
\newcommand{\br}{{\bf r}}
\newcommand{\bk}{{\bf k}}
\newcommand{\bR}{{\bf R}}
\newcommand{\bQ}{{\bf Q}}
\newcommand{\bN}{{\bf 0}}
\newcommand{\bK}{{\bf K}}
\newcommand{\bG}{{\bf G}}
\newcommand{\bb}{{\bf b}}
\newcommand{\ba}{{\bf a}}
\newcommand{\bnabla}{{\bm \nabla}}
\newcommand{\tR}{\tilde R}
\newcommand{\tV}{\tilde V_L}

\newcommand{\erf}{\mbox{erf}}
\newcommand{\erfc}{\mbox{erfc}}
\newcommand{\ds}{\displaystyle}
\newcommand{\cH}{{\cal H}}
\newcommand{\bphi}{\bm{\phi}}

\title{Planar defects and the fate of the Bragg glass phase of type-II
superconductors}

\author{T. Emig}
\affiliation{Institut f\"ur Theoretische Physik, Universit\"at zu
K\"oln, Z\"ulpicher Stra\ss e 77, 50937 K\"oln, Germany}
\affiliation{Laboratoire de Physique Th\'eorique et Mod\`eles 
Statistiques, CNRS UMR 8626, Universit\'e Paris-Sud, 91405 Orsay, France}
\author{T. Nattermann}
\affiliation{Institut f\"ur Theoretische Physik, Universit\"at zu
K\"oln, Z\"ulpicher Stra\ss e 77, 50937 K\"oln, Germany}

\date{\today}

\begin{abstract}
  It is shown that the Bragg glass phase can become unstable with
  respect to planar defects.  A single defect plane that is oriented
  parallel to the magnetic field as well as to one of the main axis of
  the Abrikosov flux line lattice is always relevant, whereas we argue
  that a plane with higher Miller index is irrelevant, even at large
  defect potentials. A finite density of parallel defects with random
  separations can be relevant even for larger Miller indices.  Defects
  that are aligned with the applied field restore locally the flux
  density oscillations which decay algebraically with distance from
  the defect.  The current voltage relation is changed to $\ln
  V(J)\sim -J^{-1}$.  The theory exhibits some similarities to the
  physics of Luttinger liquids with impurities.
\end{abstract}


\maketitle


Type-II superconductors have to contain a certain amount of disorder
to sustain superconductivity: the disorder pins magnetic flux lines,
hence preventing dissipation due to their motion
\cite{Tinkham,Blatter+93}. For some time it was believed that disorder
due to randomly distributed impurities destroys the long range
translational order (LRO) of the Abrikosov flux line lattice
\cite{Larkin}. More recently it was shown that the effect of
impurities is much weaker resulting in a phase with quasi-LRO, the
so-called Bragg glass
\cite{Nattermann90,Korshunov93,Giamarchi+94_95,Emig+99}.  In this
phase the averaged flux line density is constant but a remnant of its
periodic order is seen in the correlations of the oscillating part of
the density which decay as a power law.  Experimental signatures of
this phase have been observed \cite{bragg_glass_exp}. An important
feature of the Bragg glass is the highly non-linear
current-voltage relation related to the flux creep which is of the
form $\ln V(J)\sim -J^{-1/2}$ so that the linear resistance
vanishes.


Although much of the original transport data on flux creep in
high-T$_c$ superconductors was discussed in terms of point disorder
(see e.g. \cite{Koch+89}) it was realized later that many samples
included planar defects like twin planes or grain boundaries which
masked the Bragg glass behavior \cite{Crabtree}. Indeed, in clean
samples planar defects lead to much more pronounced pinning phenomena
than point disorder because of stronger spatial correlations
\cite{Blatter+93,Blatter+91}. However, the generic experimental
situation is a mixture of point disorder and planar defects, a case
which has not been studied theoretically in the context of the Bragg
glass yet; see however \cite{Polkovnikov+05,Nonomura+05}.




It is the aim of the present paper to consider exactly this case, i.e.
the question of the influence of planar defects in the Bragg glass
phase.  Our key results are as follows: a necessary condition for a
planar defect to become a relevant perturbation is that it is oriented
parallel to the magnetic flux. In this case its influence on the Bragg
glass phase can be characterized by the value of a \emph{single}
parameter $g\equiv\frac{3}{8}\eta(a/\ell)^2$ which depends both on the
exponent $\eta$ describing the decay of the density correlations in
the Bragg glass phase \emph{and} on the orientation of the defect with
respect to the flux line lattice. $a$ and $\ell$ are the mean spacing
of the flux lines in the bulk and the distance between lattice planes
of the Abrikosov lattice \emph{parallel} to the defect, respectively.
The defect is relevant for an infinitesimal weak defect potential if
$g<1$ which is realized if and only if the defect plane is parallel to
one of the main crystallographic planes of the flux line lattice.  In
the vicinity of the (relevant) defect the density profile shows
periodic order with an amplitude decaying algebraically (with exponent
$g$) with the normal distance from the defect.  The current voltage
relation for voltage drops perpendicular to a defect plane is of the
form $ \ln V_D(J)\sim -J^{-1}$.  Correlations are destroyed across
(relevant) defects. For $g>1$ a weak (and presumably even a strong)
defect is irrelevant and the density profile decays with a larger
exponent $2g-1>1$.  All defects which are tilted against the magnetic
flux are irrelevant and the flux density oscillations decay
exponentially. For a finite density of parallel defect planes
(with random distances) the Bragg glass is destroyed for $g<\tilde
g_c$ with $\tilde g_c\ge 3/2$.

It is worthwhile to mention that some aspects of our results are
similar to other theories at their critical dimension like
$2$-dimensional classical or $(1+1)$-dimensional quantum models.
Adding a planar defect in the Bragg glass resembles the presence of a
line defect in the classical or a frozen impurity in the quantum case
\cite{Hofstetter+04,Affleck+04,Radzhihovsky,Kane+92} when the coupling
constant $g$ is identified with temperature or the Luttinger liquid
parameter, respectively. The periodic order seen in the vicinity of
the defect plane has its counterpart in the Friedel oscillations in
Luttinger liquids close to an isolated impurity.  Whereas in those
cases the relevance of a defect (i.e. an impurity) can be changed by
tuning the interaction strength or the temperature, respectively, in
the present case a change of $g$ can be accomplished by changing the
\emph{orientation} of the defect with respect to the flux line
lattice. The current voltage relations are however different from the
2-dimensional cases.  Experimentally micro-twinned crystals with one
direction of twin planes have been studied, e.g., in
Ref. \cite{Oussena+96}.


First, we consider the effect of a single planar defect in a system
of interacting flux lines which are pinned by randomly distributed
impurities. Since we are interested in the behavior on large length
scales, it is appropriate to describe the interacting flux lines in
terms of a continuum elastic approximation with a displacement field
$\bu(\br)$. The Hamiltonian $\cH=\cH_0+\cH_D$ includes then the
  elastic energy and the impurity interaction,
\begin{eqnarray}
\label{eq:Ham} {\cH_0}&=&\frac{1}{2}\int dz d^2 \bx\left\{
c_{11}\left(\bm{\nabla}_{\bx} \cdot\bu\right)^2+
c_{66}\left(\bm{\nabla}_{\bx} \times\bu\right)^2+
\right.\nonumber \\
&+& \left. c_{44}\left(\partial_z
  \bu\right)^2  + E_{\rm pin}(\bu,\br) \right\} \, ,
\end{eqnarray}
and the coupling energy $\cH_D$ to the defect, see below. The pinning
energy can be written as $E_{\rm pin}=U(\br)\rho(\br,\bu)$, where
$U(\br)$ is a potential arising from randomly distributed impurities
and $\br \equiv (\bx,z)$. The local flux line density can be expressed
as $\rho(\br,\bu)= \rho_0\left\{- \bm{\nabla}_{\bx} \bu(\br) +
  \sum_{\bG}e^{i\bG[\bx-\bu(\br)]} \right\}$ \cite{Emig+99}. Here
$\rho_0=2/{\sqrt 3}a^2$, and $\bG\equiv{\bG_{mn}}=m \bb_1+n \bb_2$ is
a vector of the reciprocal lattice with integer $m, n$ \footnote{
  $\bb_1=\frac{2\pi}{\sqrt{3}a}(1,\sqrt{3})$,
  $\bb_2=\frac{2\pi}{\sqrt{3}a}(2,0)$, $a^2={2\Phi_0/\sqrt 3 B}$}.
The set of the six smallest reciprocal lattice vectors will be denoted
by $\{\bG_0\}$.


The energy of a planar defect {has the same form as $E_{\rm pin}$ but}
with $U(\br)$ replaced by $-v\delta(\br\cdot\bn_D-\delta)$ where $v$
is the strength of the defect potential, $\bn_D$ and $\delta$ denote
the normal vector of the defect plane and its distance (along $\bn_D$)
from the origin of the coordinate system, respectively.  This gives
\begin{equation}
\label{eq:energy_defect_1} \cH_D=\!v\rho_0 \! \int \!d^3\br \,
\delta(\br\cdot\bn_D-\delta)\big\{
  \bm{\nabla}_{\bx} \bu(\br) -\sum_{\bG\ne \bN}
e^{i\bG[\bx-\bu(\br)]}\big\} \, .
\end{equation}


Without $\cH_D$ this model has been studied in detail using different
approaches \cite{Nattermann90,Korshunov93,Giamarchi+94_95,Emig+99}.
It was shown that {thermal fluctuations are irrelevant (zero
  temperature fixed point) and that} the pinned flux line lattice
exhibits a power law decay of the translational order parameter
$\Psi_{\bG}(\bx,z)=e^{i\bG \bu(\bx,z)}$, similar to pure 2D crystals
at finite temperatures. In particular, $\langle\Psi_{\bG}\rangle=0$
but $\langle\Psi_{\bG}(\bx,0)\Psi_{{-\bG}}({\bf 0},0)\rangle\sim
|\bx|^{-\eta_{\bG}}$, where $\langle...\rangle$ denotes both the
thermal and disorder average and $\eta_{\bG}=\eta(G/G_0)^2$. From a
Gaussian variational treatment in $d=3$ dimensions follows $\eta=1$
\cite{Giamarchi+94_95} whereas a functional renormalization group
analysis in $d=4-\epsilon$ dimensions yields a non-universal exponent
$\eta$ that varies with the elastic constants of the vortex lattice
\cite{Emig+99}.  Extrapolating to $d=3$ \footnote{{A direct
    computation for $d=3$ yields a value for $\eta$ that depends not
    only on $c_{11}/c_{66}$} but on the two ratios $c_{11}/c_{44}$ and
  $c_{66}/c_{44}$. However, the range of variation of $\eta$ remains
  unchanged to order $\epsilon$.}, one finds only a very weak
variation with $1.143<\eta<1.159$ \cite{Emig+99}.  Since (despite of
the glassy nature of the phase) the structure factor shows still Bragg
peaks the notation Bragg-glass was coined \cite{Giamarchi+94_95}.
\begin{figure}
\includegraphics[width=.6\linewidth]{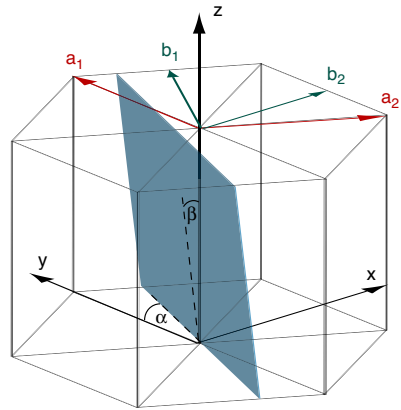}
\caption{\label{Fig:breaks}Triangular flux line lattice with
  vectors of the direct ($\ba_1$ and $\ba_2$) and the reciprocal lattice
   ($\bb_1$ and $\bb_2$) and the orientation
  of the defect plane.}
\vspace*{-0.5cm}
\end{figure}
Next we discuss the influence of $\cH_D$. In order to integrate over
the delta function in Eq.~(\ref{eq:energy_defect_1}), it is
convenient to introduce an explicit parameterization for the position
vector {$\br_D$} of the defect plane which obeys $\br_D \cdot \bn_D
=\delta $. With
\begin{eqnarray}
\label{eq:defect-para} \br_D&=&(\bx_{D},z_D)+\delta
\bn_D,\,\,\,z_D=t \cos \beta\\\nonumber \bx_{D}&=&(s \sin
\alpha - t\cos\alpha\sin\beta, s\cos\alpha+t\sin\alpha\sin\beta) 
\nonumber
\end{eqnarray}
we introduce in-plane coordinates $s$, $t$, and two angles $\alpha$
and $\beta$ which determine the rotation of the plane with respect
to the $y$- and $z$-axis, respectively ({see Figure
  \ref{Fig:breaks}}). The defect energy reads then
\begin{equation}
\label{eq:energy_defect_2} \cH_D=v \rho_0\!  \int \!  dtds
\big\{\bm{\nabla}_{\bx} \bu(\br_D) -\sum_{\bG\ne \bN}
e^{i\bG[\delta\bn_D+\bx_{D}-\bu(\br_D)]}\big\} \, .
\end{equation}
Since the displacement field $\bu(\br_D)$ varies slowly on the scale
of the flux line lattice constant $a$, the integrals over $s$ and $t
$ vanish for all $\bG$ except those for which the oscillatory factor
$e^{i\bG \bx_{D}}$ becomes {one} (for all $s$, $t$).  This condition
can be satisfied only if $\sin\beta=0$, i.e., if the defect plane is
\emph{parallel} to the applied magnetic field. There remains a
second condition for the angle $\alpha$ which results from the
constraint that $\bG$ has to be perpendicular to $\bx_{D}$.
Expressing the defect plane (for $\sin\beta=0$) as
$\bx_{D}=(c_1\ba_1-c_2\ba_2)s/a$ where $\ba_i\bb_j=2\pi\delta_{ij}$,
this results in the condition $m/n=c_2/c_1$. Hence if $c_1/c_2$ is
irrational the effect of the defect plane is always averaged to
zero. On the other hand, for rational $c_2/c_1$ we may choose
$m_D,n_D$ to be the smallest coprime pair with $c_2/c_1=m_D/n_D$.
Then $m_D$, $n_D$ are the Miller indices of the defect plane and
only those $\bG$ which are integer multiples of $\bG_D\equiv\bG_{m_D
n_D}$ contribute in Eq.~(\ref{eq:energy_defect_2}).  In the
following, we will concentrate on the contribution from
these $\bG$-vectors only. The  flux line lattice planes  (of the
ideal lattice) parallel to a  defect plane with Miller indices
$m_D$, $n_D$ have a separation of $\ell=\frac{\sqrt
3}{2}a/\sqrt{m_D^2+m_Dn_D+n_D^2}$ and hence $G_D=2\pi/\ell$.


Using the result for the average of $\Psi_\bG(\br)$ in the Bragg
  glass phase \cite{Emig+99}, one finds for the
disorder averaged defect energy $\cH_D$ on scale $L$
\begin{eqnarray}
  \label{eq:exp-averaged}
  \langle \cH_D
  \rangle_{0}\sim
  \sum_{k=1}^{\infty}v_k
\cos(k\delta G_D) \left(\frac{L}{L_a}\right)^{2-
   k^2 g},\,\,
   g\equiv\frac{3}{8}\frac{\eta a^2}{\ell^2},
\end{eqnarray}
{where $\langle\ldots\rangle_0$ denotes the average with $\cH_0$.}
The linear gradient term in (\ref{eq:energy_defect_1}) vanishes after
averaging \footnote{The coupling of $\bm{\nabla}_{\bx} \bu(\br)$ to
  the defect potential for a defect parallel to the $z$-axis can be
  eliminated by the transformation $\bu \to \bu+\bm{\nabla}_{\bx}
  \phi(\bx)$ with $\phi(\bx)$ a scalar field which obeys
  $\bm{\nabla}_{\bx}^2 \phi =
  \frac{v\rho_0}{2c_{11}}\delta(\bx\cdot\bn_D-\delta)$. This
  transformation does not change the terms $\sim c_{66}$, $c_{44}$ in
  Eq.~\eqref{eq:Ham} but shifts the flux line density on the defect.}.
The coefficients $\cos (k\delta G_D)$ reflect the periodicity of the
defect energy under translations by $\ell$ normal to the plane. The
Gaussian approximation used here is believed to be correct to order
$\epsilon$ \cite{Wiese}. It is important to remark that the
$L$-dependence of Eq.~(\ref{eq:exp-averaged}) holds only on length
scales larger than the positional correlation length $L_a\approx
L_{\xi} (a/\xi)^{1/\zeta_{rm}}$ where $\xi$ is the maximum of the
coherence and the disorder correlation length and $L_{\xi}$ denotes
the Larkin length on which the typical flux line displacement is of
the order $\xi$. $\zeta_{rm}\approx 0.175$ is the roughness exponent
of the elastic distortions on scales smaller than $L_a$ (in the
so-called random manifold regime). There the correlations of
$\Psi_{\bG}(\br)$ decay as a stretched exponential and the effect of
the defect plane is reduced by disorder fluctuations. Due to these
fluctuations on intermediate length scales the initial value of the
defect strength is reduced to $v_k\approx
v(L_a/a)^2e^{-c(G_Dka)^2}$ where $c$ is a constant.




To linear order in $v_k$, the RG flow equation of the $v_k$ {is
  obtained by comparison of the defect energy scaling in in
Eq.~(\ref{eq:exp-averaged}) with the scaling of $\cH_0$ at the Bragg
  glass fixed point,} yielding
\begin{equation}
\label{eq:RG-defect-flow}
dv_k/d\ln L=\left(1- k^2g\right)v_k \, .
\end{equation}
Hence $v_1$ is a relevant perturbation for $g<1$, i.e., if
\begin{equation}
\label{eq:relevance} \eta(m_D^2+m_Dn_D+n_D^2)<2\,\, \text{or} \quad
\ell
> \sqrt{\frac{3\eta}{8}}\,\,a\approx 0.66 \,\,a,
\end{equation}
which is compatible only with $\ell={\sqrt 3}a/2\approx 0.87 a$. Hence
the defect plane must be oriented parallel to one of the three main
crystallographic planes of the flux line lattice (i.e.
$\cos2\beta=\cos 6\alpha= 1$).




The transition described by Eq.~\eqref{eq:RG-defect-flow} occurs not
in the bulk but on the defect plane. Hence one can develop an
effective theory on the defect which could be used to describe also
stronger defect potentials. Since the defect couples only to the
normal displacement $u_{\perp}(\br_D)=\bn_D\bu(\br_D)$ on the defect
plane, we would like to integrate out $u_\perp$ outside the defect and
$\bu-u_\perp \bn_D$ across the entire sample. This integration is
facilitated by the reasonable assumption of an effective Gaussian
theory for the defect-free system at the Bragg glass fixed point
\cite{Emig+99}.  We find that $u_\perp$ on the defect has long-ranged
elasticity and is described by the effective Hamiltonian (compare
\cite{Kane+92,Radzhihovsky} for a corresponding procedure in the clean
case)
\begin{eqnarray}\label{eq:2D}
{\cal H}_\text{2D}&=&\frac{K}{2} \int d^2 \bq \, |\bq|\, |\varphi_\bq|^2
\\
& + & \int \!d^2r_D \left\{ \frac{2\sqrt{\pi g} K}{\xi} \cos(\varphi-\alpha) 
+ \frac{v_1}{L_a^2} \cos(\varphi) \right\} \, , \nonumber 
\end{eqnarray}
where $\varphi(\br_D)\equiv 2\pi u_{\perp}(\br_D)/\ell$ and $\bq$ is
the in-plane momentum. $\alpha$ is a random phase which is
uncorrelated and uniformly distributed. The amplitude of the random
potential has been chosen here as to reproduce the proper Gaussian
replica theory of the Bragg glass for $v_1=0$.  The elastic constant
$K$ depends on the bulk elastic moduli, the angle $\alpha$ and the
direction of $\bq$.  The model of Eq.~(\ref{eq:2D}) shows a transition
at $g=g_c(v_1)$ with $g_c(0)=1$ in agreement with our previous
considerations. In the present case $g= \eta(m_D^2+m_Dn_D+n_D^2)/2$
can only be changed in finite steps by changing the orientation of the
defect plane.  Thus at small $v_1$ only the defects parallel to the
three main crystallographic planes are relevant (with
$m_D^2+m_Dn_D+n_D^2=1$ ).  Due to the long-ranged elasticity we expect
from our analysis of a similar model \cite{Emig+99b} that even at
large $v_1$ one has $g_c(v_1)=1$.




Next we study the order of the flux lines in the vicinity of the
defect. Since the superconducting order is {reduced} in the defect
plane it is plausible to assume that $v>0$. The potential of a
relevant defect growth under renormalization and effectively decouples
the two half-spaces. {On large scales}, we can then impose Dirichlet
boundary conditions for the normal flux line displacement $u_\perp$ on
the defect plane. This allows us to determine the boundary induced
modifications of the flux density via the method of images.  We find
that $\langle u_\perp^2(\br)\rangle = \frac{1}{2}
\langle[u_\perp(\br)-u_\perp(\br_m)]^2\rangle_0$, where $\br_m$ is the
mirror image of $\br$ with respect to the defect plane. From this we
obtain immediately that with increasing distance $L_\perp=|\bn_D \bx -
  \delta|$ from the defect, the slowest oscillations of the flux
line density decay for a \emph{relevant} defect plane ($g<1$) as
\begin{equation}\label{eq:correlations}
{    \langle\rho(\bx,z,\bu)\rangle\sim
    \left(\frac{L_a}{L_{\perp}}\right)^g\cos(G_0(L_{\perp}\pm\delta)).}
\end{equation}
Hence, a single relevant defect plane yields a long-ranged restoration
of the order parameter. The oscillations of the density {resemble}
Friedel oscillations observed in Luttinger liquids close to an
isolated impurity \cite{Kane+92}. This similarity is substantiated by
considering the decay of the density oscillations if the defect plane
is \emph{irrelevant}, i.e., for $g>1$. Then the defect potential
decreases under renormalization and lowest order perturbation theory
can be applied. Such an approach takes into account that the defect
strength decreases as $v(L)\sim L^{1-g}$ and hence we obtain
Eq.~(\ref{eq:correlations}) with $g$ replaced by $2g-1>1$, in close
analogy to the ($1+1$)-dimensional counterpart
\cite{Hofstetter+04,Affleck+04,Radzhihovsky}.




If the defect plane is {\it not} parallel to the applied magnetic
field, i.e. $\sin\beta \neq 0$, then the irrelevance of the defect
should be also observable in the Friedel oscillations of the flux line
density which should become then short-ranged.  If $g>1$ we can use
perturbation theory to obtain the density oscillations since the
defect strength deceases for large length scales, i.e., for
sufficiently small $\sin\beta$. From this analysis we find for the
slowest oscillations normal to the defect (with $\delta=0$)
\begin{equation}
\label{eq:osci+tilt} \langle\rho(\bx,z,\bu)\rangle  \sim
\left(\frac{L_a^2}{\xi L_\perp} \right)^{g-\frac{1}{2}}
e^{-L_\perp/\xi} \cos\left(G_D L_\perp \cos\beta\right)
\end{equation}
for large separations $L_\perp \gg \xi$ with a characteristic length
scale $\xi=1/(G_D|\sin\beta|)$. Hence, if the tilt angle $\beta$
approaches zero, $\xi$ diverges, and the Friedel oscillations for
$L_\perp \ll \xi$ are described by Eq.(\ref{eq:osci+tilt}) with
$\xi$ replaced by $L_\perp$, hence in agreement with our findings
for $\beta=0$. Although our perturbative analysis is restricted to
$g>1$ we expect that for any value of $g$ the Friedel oscillations
decay exponentially beyond distances of order $\xi$.  Since the
renormalization of the defect potential is cutoff on the scale
$\xi$, even for a relevant defect perturbation theory is expected to
be justified if both the bare defect potential is sufficiently weak
and $\sin\beta$ is not too small.




To find the current voltage relation for a relevant defect we consider
the most interesting case where the current $J$ is parallel to the
defect and normal to the magnetic field $\bf B$. Flux creep in the
presence of pinning forces arises via formation of critical droplets
\cite{Blatter+93}. In the present case of a defect plane the droplet
is characterized by $G_Du_{\perp}(\br_D)=0$ and
$2\pi$ outside and inside of the droplet, respectively.  Since on scale
$L_a$ typical distortions in the Bragg glass phase are of the order
$a$, the droplet volume is $\sim L^2L_a$ and the volume energy gain in
the droplet is of the order $\sim -JBL^2L_a$ which has to be compared
with the the surface energy loss of the order $\sim
v_1^{1/2(1-g)}LL_a^{1/2}$ where we have included the renormalization
of $v_1$.  From the balance of the two terms follows a critical
droplet size {$L_J\sim v_1^{1/2(1-g)}(L_aJB)^{-1}$}.  Droplets of this
size have a free energy $\sim v_1^{1/(1-g)}(JB)^{-1}$ which determines
via the Arrhenius law the creep velocity due to thermal activation,
yielding the voltage drop normal to the defect plane
\begin{equation}\label{}
    V\sim \exp\left[-(C_0/(TJ))\right] ,\,\,\,\,\,\,\,
C_0\sim v_1^{1/(1-g)}B^{-1}.
\end{equation}
  Thus the flux creep across the defect is much
slower than in the Bragg glass phase. {If for weak defects $g \to
  1-$, $C_0$ becomes small, hence} reducing the applicability of this
formula to extremely small currents.


Crystals usually contain either two orthogonal families (``colonies'')
of parallel twin boundaries or a {\it single} family of parallel twin
planes. We focus on the latter case, which for impure samples has been
studied so far only in (1+1) dimensions \cite{Hwa+93}. We consider the
typical situation of planes with random distances $\delta_j$ that are
aligned with the applied field. Each plane contributes
$\cH_D(\delta_j)$ of Eq.~\eqref{eq:energy_defect_2} with $\delta_j$
replacing $\delta$.  For the averages over the
$\delta_j$ one finds $\overline{\langle \sum_j \cH_D(\delta_j)
  \rangle}=0$ but $\overline{\langle \sum_j \cH_D^2(\delta_j)
  \rangle}^{1/2}\sim L^{5/2-g}$ from the term with $k=1$.  Since
$\cH_0\sim L$, a family of weak defect planes is relevant and destroys
the Bragg glass for $g<\tilde g_c=3/2$, i.e., the planes must be
oriented with a main plane of the flux lattice.  Contrary to a single
defect plane, we expect that a larger defect strength yields an
increased $\tilde g_c>3/2$ rendering additional defect orientations
relevant.  However, a description of the transition at strong coupling
and the localized fixed point describing relevant defect planes is not
available at present \cite{Balents93}.  Only for defect planes of
equal distance we expect long range order in the direction
perpendicular to the planes.

We acknowledge useful comments by L. Radzihovsky, E. Zeldov, and in
particular by D.~R.~Nelson who brought the problem to our attention,
and TN thanks R. Woerdenweber for conversations.  This work was
supported by DFG through Emmy Noether grant EM70/2 (TE), and SFB 608
(TE and TN).

\end{document}